\documentclass[aps,prb,twocolumn,floatfix,showpacs,showkeys]{revtex4}

\begin{document}

\title{Translation-symmetry protected topological orders on lattice}

\author{Su-Peng Kou}
\affiliation{Department of Physics, Beijing Normal University,
Beijing, 100875 P. R. China }

\author{Xiao-Gang Wen}
\homepage{http://dao.mit.edu/~wen}
\affiliation{Department of
Physics, Massachusetts Institute of Technology, Cambridge,
Massachusetts 02139}
\affiliation{Perimeter Institute for
Theoretical Physics, 31 Caroline Street North, Waterloo, Ontario N2J
2Y5, Canada}

\begin{abstract}
In this paper we systematically  study a simple class of
translation-symmetry protected topological orders in quantum spin
systems using slave-particle approach. The spin systems on square
lattice are translation invariant, but may break any other symmetries.
We consider topologically ordered ground states that do not
spontaneously break any symmetry.  Those states can be described by
Z2A or Z2B projective symmetry group.  We find that the Z2A
translation symmetric topological orders can still be divided into 16
sub-classes corresponding to 16 new translation-symmetry protected
topological orders.  We introduced four $Z_2$ topological indices
$\zeta_{\v k}=0,1$ at $\v {k}=(0,0)$, $(0,\pi )$ , $(\pi ,0)$, $(\pi
,\pi )$ to characterize those 16 new topological orders.  We
calculated the topological degeneracies and crystal momenta for those
16 topological phases on even-by-even, even-by-odd, odd-by-even, and
odd-by-odd lattices, which allows us to physically measure such
topological orders.  We predict the appearance of gapless fermionic
excitations at the quantum phase transitions between those symmetry
protected topological orders.  Our result can be generalized to any
dimensions.  We find 256 translation-symmetry protected Z2A
topological orders for a system on 3D lattice.
\end{abstract}

\pacs{75.45.+j, 71.27.+a}
\keywords{topological order, topological degeneracy, spin liquid}

\maketitle

\section{Introduction and Motivation}

For long time, people believe that Landau symmetry breaking
theory\cite{L3726} and the associated local order
parameters\cite{GL5064,LanL58} describe all kinds of phases and
phase transitions.  However, in last 20 years, it became more and
more clear that Landau theory cannot describe all quantum states of
matter (the states of matter at zero temperature).\cite{Wtop,Wrig} A
nontrivial example of new states of matter beyond Laudau theory is
the fractional quantum hall (FQH) states.\cite{TSG8259} FQH states
do not break any symmetry and hence cannot be described by broken
symmetries. The subtle structures that distinguish different FQH
states are called topological order.\cite{Wtop,Wrig,Wtoprev}
Physically, topological order describes the internal order (or more
precisely, the long range entanglement) in a gapped quantum ground
state. It can be (partially) characterized by robust ground-state
degeneracy.\cite{Wtop,Wsrvb} Recently, many different systems with
topologically ordered ground states were
found.\cite{KL8795,WWZcsp,RS9173,Wsrvb,K032,MS0181,BFG0212,MSP0202,Wqoexct}

Quantum spin liquid states in general contain non-trivial
topological orders. In the projective construction of spin liquid
(also known as slave-particle
approach),\cite{BZA8773,BA8880,DFM8826,AZH8845,WL9603,SF0050,Wen04}
there exist many spin liquids with low energy $SU(2)$, $U(1)$ or
$Z_2$ gauge structures. Those spin liquids all have the exactly the
same symmetry.  To distinguish those spin liquids, we note that
although the spin liquids have the same symmetry, within the
projective construction, their ansatz are not directly invariant
under the translations.  The ansatz are invariant under the
translations followed by different gauge transformations. So the
invariant group of the ansatz are different.  We can use the
invariant group of the ansatz to characterize the new order in the
spin liquids.  The invariant group of an ansatz is formed by all the
combined symmetry transformations and the gauge transformations that
leave the ansatz invariant. Such a group is called the Projective
Symmetry Group (PSG).\cite{Wqoslpub} Thus although one cannot use
symmetry and order parameter to describe different  orders in the
spin liquids, one can use the PSG to characterize/distinguish the
different quantum/topological orders of spin liquid states.

The simplest kind of topological orders is the $Z_2$ topological order
where the slave-particle ansatz is invariant under a $Z_2$ gauge
transformation.
According to the PSG characterization within the projective
construction, for system with only lattice translation symmetry,
there can be two different classes of $Z_2$ topological orders
labeled by Z2A and Z2B (a Z2B ansatz has $\pi$ flux going through
each plaquette).\cite{Wqoslpub} In this paper, we will study the Z2A
topological orders and ask ``are there distinct Z2A topological
orders?'' We find that there are indeed distinct Z2A topological
orders. They can be labeled by four $Z_2$ topological indices
$\zeta_{\v k}=0,1$ at $\v {k}=(0,0)$, $(0,\pi )$ , $(\pi ,0)$, $(\pi
,\pi )$.  So the $\zeta_{\v k}$ characterization is beyond the PSG
characterization of quantum/topological order and provides
additional information for translation-symmetry protected $Z_2$
topological order.

\section{A general ``mean-field'' fermion Hamiltonian of $Z_2$ topological
orders}
\label{gmean}

We will use the projective construction (the slave-particle
theory)\cite{BZA8773,Wen04} to systematically construct different
translation symmetric $Z_2$ topological orders in spin-1/2 systems on
square lattice. In such a construction, we start with ``mean-field''
fermion Hamiltonian\cite{Wqoslpub}
\begin{equation}
H_\text{mean}= \sum_{\v i\v j} \psi_{\v i}^{\dag }u_{\v i\v
j}\psi_{\v j} +\sum_{\v i\v j}(\psi_{\v i}^{\dag }\eta_{\v i\v
j}\psi_{\v j}^{\dag }+h.c.) +\sum_{\v i}\psi_{\v i}^{\dag }a_{\v
i}\psi_{\v i} \label{mean}
\end{equation}
where $u_{\v i\v j}$, $\eta_{\v i\v j}$, $a_{\v i}$ are 2 by 2
complex matrices.  The $\eta$-term is included since our spin-1/2
systems in general do not have any spin rotation symmetry. We like to
mention that $a_{\v i}$ are not free parameters. $a_{\v i}$ should be
chosen such that
\begin{equation}
\langle \Psi_\text{mean}^{(u_{\v i\v j},\eta_{\v i\v j},a_{\v i})}|\psi_{\v
i}^{\dagger
}\sigma ^l\psi_{\v i}|\Psi_\text{mean}^{(u_{\v i\v j},\eta_{\v
i\v j},a_{\v
i})}\rangle
=0,\ \ \ \ \ l=1,2,3
\end{equation}
where $\sigma ^l$ are the Pauli matrices.  In this paper, we will
only consider translation invariant ansatz $u_{\v i\v j}=u_{\v i+\v
a,\v j+\v a}$ and $\eta_{\v i\v j}=\eta_{\v i+\v a,\v j+\v a}$.
Those states are characterized by Z2A PSG and are Z2A topological
states.\cite{Wqoslpub}

Let $|\Psi_\text{mean}^{(u_{\v i\v j},\eta_{\v i\v j})}\rangle $
be the ground state of $H_\text{mean}$. Then a many-spin state can
be obtained from the mean-field state $|\Psi_\text{mean}^{(u_{\v
i\v j},\eta_{\v i\v j})}\rangle $ by projection
\begin{equation}
|\Psi_{\text{spin}}^{(u_{\v i\v j},\eta_{\v i\v j})}\rangle =\mathcal{P}|\Psi_{
\text{mean}}^{(u_{\v i\v j},\eta_{\v i\v j})}\rangle
\end{equation}
into the subspace with even numbers of fermion per site.
Here the projection operator is
\[
\mathcal{P}=\prod_{\v i}\frac{1+(-1)^{n_{\v i}}}2,
\]
and $n_{\v i}=\psi_{\v i}^{\dag }\psi_{\v i}$ is fermion operator at site $i$.

We note that after the projection, each site can have either no
fermion or two fermions. If we associate the no-fermion state as the
spin-down state and the two-fermion state as the spin-up state, then
the projected state $ |\Psi_{\text{spin}}^{(u_{\v i\v j},\eta_{\v
i\v j})}\rangle $ can be viewed as a quantum state for the spin
system. This is how we construct many-spin state from the mean-field
Hamiltonian. For each choice of the ansatz $(u_{\v i\v j},\eta_{\v
i\v j},a_{\v i})$, this procedure produces a physical many-spin wave
function $ |\Psi_{\text{spin}}^{(u_{\v i\v j},\eta_{\v i\v
j})}\rangle $. So the ansatz $ (u_{\v i\v j},\eta_{\v i\v j})$ can
also be viewed as a set of labels that label a many-spin state and
$|\Psi_{\text{spin}}^{(u_{\v i\v j},\eta_{\v i\v j})}\rangle$ can
be view as a trial wave function for a spin-1/2 system, with $u_{\v
i\v j}$ and $\eta_{\v i\v j}$ being variational parameters.

For the translation invariant ansatz, one can rewrite the ``mean-field''
fermion Hamiltonian in momentum space by introducing
\[
\Psi_{\v {k}}=\left(
\begin{array}{l}
\psi_{1,\v {k}} \\
\psi_{1,-\v {k}}^{\dag } \\
\psi_{2,\v {k}} \\
\psi_{2,-\v {k}}^{\dag }
\end{array}
\right)
\]
and
\[
\Psi_{\v {k}}^{\dag }=\left(
\begin{array}{llll}
\psi_{1,\v {k}}^{\dag } & \psi_{1,-\v {k}} & \psi_{2,\v {k}
}^{\dag } & \psi_{2,-\v {k}}
\end{array}
\right) .
\]
Note that $\Psi_{\v {k}}$ satisfy the following algebra
\[
\{\Psi_{I\v {k}}^{\dag },\Psi_{J\v {k}^{\prime }}\}=\delta
_{IJ}\delta_{\v {k}-\v {k}^{\prime }},\ \ \ \ \ \ \{\Psi_{I\v {
k}},\Psi_{J\v {k}^{\prime }}\}=\Ga_{IJ}\delta_{\v {k}+\v {k}
^{\prime }}
\]
where
\begin{equation}
\Ga=
\sigma_1\otimes \sigma_0
=
\bpm
\sigma_1 & 0\\
0 & \sigma_1\\
\epm
.
\end{equation}
We also note that $(\Psi_{-\v {k}
}^{\dag },\Psi_{-\v {k}})$ can be expressed in term of $(\Psi_{\v {k}
}^{\dag },\Psi_{\v {k}})$:
\begin{equation}
\Psi_{-\v {k}}=\Gamma \Psi_{\v {k}}^{*},\ \ \ \ \ \ \ \ \Psi_{-
\v {k}}^{\dag }=\Psi_{\v {k}}^T\Gamma .
 \label{Psikmk}
\end{equation}

In terms of $\Psi_{\v {k}}$, $H_\text{mean}$ can be written as
\begin{equation}
H_\text{mean}=
\sum_{\v {k}\neq 0}\Psi_{\v {k}}^{\dag }M(\v {k})\Psi_{\v {k}}
+\sum_{\v {k} = 0}\Psi_{\v {k}}^{\dag }M(\v {k})\Psi_{\v {k}}
\label{Hmean1}
\end{equation}
where $-\pi < k_x,k_y <+\pi$ and $M(\v {k})$ are $4\times 4$
Hermitian matrices $M(\v {k})=M^\dag(\v {k})$. Here $\v {k}=0$ means
that $(k_x,k_y)=(0,0)$, $(0,\pi )$, $(\pi ,0)$, or $(\pi ,\pi )$.
Also $k_x$ and $k_y$ are quantized: $k_x=\frac{2\pi}{L_x}\times$
integer and $k_y=\frac{2\pi}{L_y}\times$ integer, where $L_x$ and
$L_y$ are size of the square lattice in the $x$- and $y$-directions.
Note that on an even by even lattice (\ie $L_x=$ even and $L_y=$
even), $(k_x,k_y)=(0,0)$, $(0,\pi )$, $(\pi ,0)$, or $(\pi ,\pi )$
all satisfy the quantization conditions $k_x=\frac{2\pi}{L_x}\times$
integer and $k_y=\frac{2\pi}{L_y}\times$ integer.  In this case,
$\sum_{\v {k} = 0}$ sums over all the four points $(k_x,k_y)=(0,0)$,
$(0,\pi )$, $(\pi ,0)$, and $(\pi ,\pi )$. On other lattices,
$\sum_{\v {k} = 0}$ sums over less points.  Say on an odd by odd
lattice, only $(k_x,k_y)=(0,0)$ satisfies the quantization
conditions $k_x=\frac{2\pi}{L_x}\times$ integer and
$k_y=\frac{2\pi}{L_y}\times$ integer.  In this case,  $\sum_{\v {k}
= 0}$ sums over only $(k_x,k_y)=(0,0)$ point.

We note that
\[
\Psi_{\v {k}}^{\dag }\Psi_{\v {k}}=2,\ \ \ \ \ \ \Psi_{\v {k}
}^{\dag }\sigma_0\otimes \sigma_3\Psi_{\v {k}}=0.
\]
Thus up to a constant in $H_\text{mean}$, we may assume $M(\v {k})$
to satisfy
\begin{equation}
\text{Tr}M(\v {k})=0,\ \ \ \ \ \ \ \text{Tr}[M(\v {k})\sigma
_0\otimes \sigma_3]=0 \label{TrM} .
\end{equation}
Due to Eq. (\ref{Psikmk}),
\[
\Psi_{-\v {k}}^{\dag }M(-\v {k})\Psi_{-\v {k}}=\text{Tr}M( \v
{k})-\Psi_{\v {k}}^{\dag }\Gamma M^T(-\v {k})\Gamma \Psi_{ \v {k}}.
\]
Thus, we may rewrite Eq. (\ref{Hmean1}) as
\begin{align}
\label{meanU}
H_\text{mean} &=
\sum_{\v {k}>0}\Psi_{\v {k}}^{\dag }U( \v {k})\Psi_{\v {k}}
+\frac12 \sum_{\v {k}=0}\Psi_{\v {k}}^{\dag }U( \v {k})\Psi_{\v {k}}
,
\nonumber\\
U(\v k)&=M(\v k) -\Ga M^T(-\v k) \Ga .
\end{align}
Here
$\v {k}>0$ means that $\v {k}\neq 0$ and $k_y>0$ or $k_y=0,\ k_x>0$.
Clearly $U( \v {k})$ satisfy
\[
U(\v {k})=-\Gamma U^T(\v {-k})\Gamma ,\ \ \ \ \ \
U(\v {k})=U^\dag(\v {k}) .
\]

Now we expand $U(\v {k})$ by $16$ Hermitian matrices
\begin{equation}
 M_{\{\alpha \beta \}} \equiv
\si_\al\otimes \si_\bt, \ \ \ \
\al,\bt =0,1,2,3 ,
\end{equation}
where $\si_0=\v 1$.  We have
\[
U(\v {k})=\sum_{\{\alpha ,\beta \}}c_{\{\alpha \beta \}}(\v {k}
)M_{\{\alpha \beta \}}
\]
where $c_{\{\alpha \beta \}}(\v {k})$ are real.  The $16$ $4\times 4$
matrices $M_{\{\al\beta \}}$ can be divided into two classes :
in one class, the matrices satisfy
\[
M=-\Gamma M^T\Gamma .
\]
We call them ``\textit{even
matrices}'';
In the other class, the matrices satisfy
\[
M=\Gamma M^T\Gamma .
\]
We call them ``\textit{odd matrices}''.

There are six even matrices,
\begin{eqnarray*}
M_{\{ 30\} } &=&\sigma_3\otimes \si_0, \text{ }M_{\{
12\} }=\sigma_1\otimes \sigma_2,\text{ }M_{\{ 22\}
}=\sigma_2\otimes \sigma_2, \\
M_{\{ 33\} } &=&\sigma_3\otimes \sigma_3\v {,}\text{ }
M_{\{ 31\} }=\sigma_3\otimes \sigma_1,\text{ }M_{\{
02\} }=\si_0\otimes \sigma_2.
\end{eqnarray*}
For above six matrices, $M_{\{ 30\} },$ $M_{\{ 12\} }$
and $M_{\{ 22\} }$ anti-commute with each other,
\begin{align}
\{ M_{\{ 30\} },M_{\{ 12\} }\} &=0,
\nonumber\\
\{ M_{\{ 30\} },M_{\{ 22\} }\} &=0,
\nonumber\\
\{ M_{\{ 12\} },M_{\{ 22\} }\} &=0.
\end{align}
$M_{\{ 33\} },$ $M_{\{ 31\} }$and $M_{\{ 02\} }$ anti-commute each
other
\begin{align}
\{ M_{\{ 33\} },M_{\{ 31\} }\} &=0,
\nonumber\\
\{ M_{\{ 33\} },M_{\{ 02\} }\} &=0,
\nonumber\\
\{ M_{\{ 31\} },M_{\{ 02\} }\} &=0.
\end{align}
While each of $M_{\{ 30\} },$ $M_{\{ 12\} }$ and $
M_{\{ 22\} }$ commutate with each of $M_{\{ 33\} },
M_{\{ 31\} }$ and $M_{\{ 02\} }$. For the coefficients
of the even matrices $c_{\{\alpha \beta \}}(\v {k})$ in the
``mean-field'' fermion Hamiltonian, $\{\al\beta \}=\{ 30\}
,\{ 12\} ,\{ 22\} ,\{ 33\} ,\{
31\} ,\{ 02\} $, we have
\[
c_{\{\alpha \beta \}}(\v {k})=c_{\{\alpha \beta \}}(-\v {k}).
\]

In addition, there exist ten odd matrices:
\begin{align*}
M_{\{ 00\} } &=\si_0\otimes \si_0,
&
M_{\{ 10\} }&=\sigma_1\otimes \si_0,
&
M_{\{ 20\} }&=\sigma_2\otimes \si_0,
\\
M_{\{ 03\} } &=\si_0\otimes \sigma_3,
&
M_{\{ 11\} }&=\sigma_1\otimes \sigma_1,
&
M_{\{ 13\} } &=\sigma_1\otimes \sigma_3,
\\
M_{\{ 23\} } &=\sigma_2\otimes \sigma_3,
&
M_{\{ 32\} }&=\sigma_3\otimes \sigma_2,
&
M_{\{ 21\} }&=\sigma_2\otimes \sigma_1,
\\
M_{\{ 01\} }&=\si_0 \otimes \sigma_1 .
\end{align*}
For the coefficients of odd matrices $c_{\{\alpha \beta \}}(\v {k})$ in
the ``mean-field'' fermion Hamiltonian, we have
\[
c_{\{\alpha \beta \}}(\v {k})=-c_{\{\alpha \beta \}}(-\v {k}) .
\]
Thus for odd matrices, $
c_{\{\alpha \beta \}}(\v {k})$ are odd functions of $k_x,k_y$ and are
fixed to be zero at momentum $(0,0)$, $(0,\pi )$, $(\pi ,0)$, $(\pi ,\pi )$
\[
c_{\{\alpha \beta \}}(\v {k}=0) = 0.
\]

\section{Classification of Z2A topological orders}

For a generic choice of $u_{\v i\v j}$ and $\eta_{\v i\v j}$, the
corresponding mean-field Hamiltonian \eq{meanU} is gapped.  Note
that the energy levels of the mean-field Hamiltonian \eq{meanU}
appear in $(E,-E)$ pairs. The mean field state is obtained by
filling all the negative energy levels.  The mean-field Hamiltonian
is gapped if the minimal positive energy level is finite.  The
gapped mean-field Hamiltonian corresponds to a gapped Z2A spin
liquid.

As we change the mean-field parameters $u_{\v i\v j}$ and $\eta_{\v
i\v j}$, the mean-field energy gap (and the correspond energy gap
for the Z2A spin liquid) may close which indicates a quantum phase
transition.  Thus if two gapped regions are always separated by a
gapless region, then the two gapped regions will correspond to two
different quantum phases. We may say that the two quantum phases
carry different topological orders.

In the following, we introduce topological indices that can be
calculated for each gapped mean-field ansatz $(u_{\v i\v j},\eta_{\v
i\v j})$.  We will show that two gapped mean-field ansatz with
different topological indices cannot smoothly deform into each other
without closing the energy gap.  Therefore, the topological indices
characterize different Z2A topological orders with translation
symmetry.

\subsection{The topological indices}

In the section \ref{gmean}, we obtained the ``mean-field'' fermion
Hamiltonian in momentum space \eq{meanU},
which has a form
$H_\text{mean}=H(\v{k}>0)_\text{mean}
+H( \v{k}=\v{0})_\text{mean}$.
Let us diagonalizing the ``mean-field'' fermion Hamiltonian at the points
$\v{k}>0$ as an example. Introducing
\begin{align}
W(\v k) \Psi_{\v{k}}&=\left(
\begin{array}{l}
\alpha_{\v{k}} \\
\alpha_{-\v{k}}^{\dag } \\
\beta_{\v{k}} \\
\beta_{-\v{k}}^{\dag }
\end{array}
\right)
,
\end{align}
where
\begin{align}
&\ \ \ W(\v k)U(\v{k})W^\dag(\v k)
\nonumber\\
&=\left(
\begin{array}{llll}
\varepsilon_1(\v{k}) & 0 & 0 & 0 \\
0 & -\varepsilon_1(\v{k}) & 0 & 0 \\
0 & 0 & \varepsilon_2(\v{k}) & 0 \\
0 & 0 & 0 & -\varepsilon_2(\v{k})
\end{array}
\right) ,
\end{align}
$\varepsilon_1(\v{k})>0$, and $\varepsilon_2(\v{k})>0$,
we find
\begin{widetext}
\begin{eqnarray*}
H(\v{k} >0)_\text{mean}=\sum_{\v{k}>0}
\Big[\varepsilon_1(\v{k})\alpha_{\v{k}}^{\dag }\alpha_{\v{k}}
+\varepsilon_2(\v{k})\beta_{\v{k}}^{\dag }\beta_{\v{k}}
-\varepsilon_1(\v{k})\alpha_{-\v{k}}\alpha_{-\v{k}}^{\dagger }
-\varepsilon_2(\v{k})\beta_{- \v{k}}\beta_{-\v{k}}^{\dag } \Big].
\end{eqnarray*}
We note that both $\alpha_{\pm \v{k}}$ and $\beta_{\pm \v{k}}$
will annihilate the mean-field ground state $|\Psi_{\text{\text{mean}}%
}\rangle ,$
\[
\alpha_{\pm \v{k}}|\Psi_\text{mean}\rangle =0,\ \ \ \ \ \
\ \ \beta_{\pm \v{k}}|\Psi_\text{mean}\rangle =0.
\]

At the four $\v k=0$ points, only the even $M_{\al,\bt}$ appear and
we diagonalized the Hamiltonian differently. The four eigenvalues of
$U(\v k)$ are given by
\begin{align}
&
\veps_{\pm\pm}(\v k)=\pm \sqrt{c_{\{30\}}^2(\v {k} )+c_{\{12\}}^2(\v {k})
+c_{\{22\}}^2(\v {k})}
 \pm\sqrt{c_{\{33\}}^2(\v {k})+c_{\{31\}}^2(\v {k})
+c_{\{02\}}^2( \v {k})}
\end{align}
and
\begin{eqnarray*}
H(\v{k} =0)_\text{mean}
&=&\frac12 \sum_{\v{k}=0}\Big[
\varepsilon_{++}(\v k)\alpha_{\v{k}}^{\dag }\alpha_{\v{k}}
+\varepsilon_{+-}(\v{k})\beta_{\v{k}}^{\dag }\beta_{\v{k}}
+\varepsilon_{--}(\v{k})\alpha_{\v{k}}\alpha_{\v{k}}^{\dagger}
+\varepsilon_{-+}(\v{k})\beta_{\v{k}}\beta_{\v{k}}^{\dag }
\Big]
\nonumber\\
&=& \sum_{\v{k}=0}\Big[
\varepsilon_{++}(\v k)\alpha_{\v{k}}^{\dag }\alpha_{\v{k}}
+\varepsilon_{+-}(\v{k})\beta_{\v{k}}^{\dag }\beta_{\v{k}}
\Big]+\text{Const.}
\end{eqnarray*}
\end{widetext}
where
\begin{align}
&\ \ \  W(\v k)U(\v{k})W^\dag(\v k)
\nonumber\\
&=\left(
\begin{array}{llll}
\varepsilon_{++}(\v{k}) & 0 & 0 & 0 \\
0 & -\varepsilon_{++}(\v{k}) & 0 & 0 \\
0 & 0 & \varepsilon_{+-}(\v{k}) & 0 \\
0 & 0 & 0 & -\varepsilon_{+-}(\v{k})
\end{array}
\right) .
\end{align}
We note that $W(\v k)$ diagonalizes the linear combination of
$M_{\{30\}}$, $M_{\{12\}}$, and $M_{\{22\}}$ in
$H_\text{mean}$:
\begin{align}
&\ \ \ W \left (c_{\{30\}}M_{\{30\}}+c_{\{12\}}M_{\{12\}}+c_{\{22\}}M_{\{22\}}
\right )W^\dag
\nonumber\\
&= \sqrt{c_{\{30\}}^2+c_{\{12\}}^2+c_{\{22\}}^2} M_{\{30\}}.
\end{align}
$W(\v k)$ also diagonalizes the linear combination of
$M_{\{33\}}$, $M_{\{31\}}$, and $M_{\{02\}}$ in
$H_\text{mean}$:
\begin{align}
&\ \ \ W \left (c_{\{33\}}M_{\{33\}}+c_{\{31\}}M_{\{31\}}+c_{\{02\}} M_{\{02\}}
\right )W^\dag
\nonumber\\
&= \sqrt{c_{\{33\}}^2+c_{\{31\}}^2+c_{\{02\}}^2} M_{\{33\}}.
\end{align}
So $W(\v k)$ changes $M_{33}=\si_3\otimes \si_3$ to
\begin{align}
\label{W33W}
&\ \ \ W(\v k) (\si_3\otimes \si_3) W^\dag(\v k)
\nonumber\\
&=
a(\v k) \si_3\otimes \si_3
+b(\v k) \si_0\otimes \si_1
+c(\v k) \si_3\otimes \si_2 .
\end{align}
where $a^2(\v k)+b^2(\v k)+c^2(\v k)=1$.

The energy spectrum at $\v k=0$ motivates
us to introduce $\zeta_{\v k}$ as
the four topological indices, one for each $\v k=0$ point:
\begin{align}
\zeta_{\v k} &=1-\Th [\veps_{+-}(\v k)] .
\\
&=
1-\Th[
c_{\{30\}}^2(\v {k} )+c_{\{12\}}^2(\v {k})
+c_{\{22\}}^2(\v {k})
\nonumber\\
&\ \ \ \ \ \ \ \ \ \
- c_{\{33\}}^2(\v {k})-c_{\{31\}}^2(\v {k})
-c_{\{02\}}^2( \v {k})
]
\nonumber
\end{align}
where $\Th(x)=1$ if $x>0$ and $\Th(x)=0$ if $x<0$.  If two
topological ordered states have different sets of topological
indices $(\zeta_{\v k=(0,0)}, \zeta_{\v k=(\pi,0)}, \zeta_{\v
k=(0,\pi)}, \zeta_{\v k=(\pi,\pi)})$, then as we deform one state
smoothly into the other, some $\zeta_{\v k}$ must change. $\zeta_{\v
k}$ can only change when $ \sqrt{c_{\{30\}}^2(\v {k}
)+c_{\{12\}}^2(\v {k}) +c_{\{22\}}^2(\v {k})} =\sqrt{c_{\{33\}}^2(\v
{k})+c_{\{31\}}^2(\v {k}) +c_{\{02\}}^2( \v {k})} $. At that point,
the topological ordered state becomes gapless indicating a quantum
phase transition. Therefore, there are 16 different translation
invariant Z2A spin liquids labeled by $\zeta_{\v k=(0,0)}$,
$\zeta_{\v k=(0,\pi )}$, $\zeta_{\v k=(\pi,0)}$,
$\zeta_{\v k=(\pi ,\pi )} =$ 1111, 1100,
1010, 1001, 0101, 0011, 0110, 0000, 1000, 0100, 0010, 0001, 1110,
1101, 1011,  0111.

\subsection{The topological degeneracy}

Let's calculate topological degeneracies for different topological
orders through the projected construction. Now we use $|m,n\rangle
=|\Psi_\text{mean}^{(u_{\v i\v j}^{(m,n)},\eta_{\v i\v
j}^{(m,n)})}\rangle $ $(|m,n\rangle =|0,0\rangle ,$ $|0,1\rangle ,$
$|1,0\rangle ,$ $|1,1\rangle )$ to note four degenerate ground states
on a torus. Here $(u_{\v i\v j}^{(m,n)},\eta_{\v i\v j}^{(m,n)})$ is
defined as $ ((-)^{ms_x(\v i\v j)}(-)^{ns_y(\v i\v j)}u_{\v i\v
j},(-)^{ms_x(\v i\v j)}(-)^{ns_y(\v i\v j)}\eta_{\v i\v j}).$
$s_{x,y}(\v i\v j)$ have values $0$ or $1$, with $s_{x,y}(\v i\v j)=1$
if the link $\v i\v j$ crosses the $x=L_x$, or $y=L_y$ line(s) and
$s_{x,y}(\v i\v j)=0$ otherwise\cite{Wqoslpub}.  In fact, the four
mean-field states $|m,n\rangle =|\Psi_\text{mean}^{(u_{\v i\v
j}^{(m,n)},\eta_{\v i\v j}^{(m,n)})}\rangle $ are obtained by giving
the fermion wave-functions $\psi (x,y)$ different boundary
conditions:
\begin{align}
\psi (x, y)&=(-1)^m\psi (x,\text{ }y+L_y),
\nonumber\\
\psi (x, y)&=(-1)^n\psi (x+L_x,\text{ }y).
\end{align}

To obtain a physical state from the mean-field ansatz
$|\Psi_\text{mean}^{(u_{\v i\v j},\eta_{\v i\v j})}\rangle$, one
needs to project $|\Psi_\text{mean}^{(u_{\v i\v j},\eta_{\v i\v
j})}\rangle $ into the subspace with even numbers of $\psi$ fermion
per site.  So the mean-field state must have even numbers of $\psi$
fermions in order for the projection to be non-zero.  The total $\psi$
fermion number has a form $\hat{N}=N_{\v k\neq 0}+N_{\v k= 0}$ where
$N_{\v {k}\neq \v {0}}=\sum\limits_{ \v {k>0}} (\psi_{\v
k}^\dag\psi_{\v k}+\psi_{-\v k}^\dag\psi_{-\v k})$ and $N_{\v {k}
=0}=\sum\limits_{\v {k}=0}\psi_{\v {k}}^{\dag }\psi_{ \v {k}}$.

We note that, for $\v {k}>0$,
\begin{eqnarray}
&&
(-)^{
\psi_{1,\v {k}}^{\dag }\psi_{1,\v {k}}
+\psi_{1,-\v {k}}^{\dag }\psi_{1,-\v {k}}
+\psi_{2,\v {k}}^{\dag }\psi_{2,\v {k }}
+\psi_{2,-\v {k}}^{\dag }\psi_{2,-\v {k}}
}
\nonumber\\
&=&
(-)^{
\psi_{1,\v {k}}^{\dag }\psi_{1,\v {k}}
-\psi_{1,-\v {k}}\psi_{1,-\v {k}}^\dag
+\psi_{2,\v {k}}^{\dag }\psi_{2,\v {k }}
-\psi_{2,-\v {k}}\psi_{2,-\v {k}}^\dag
}
\nonumber \\
&=&
(-)^{
\psi_{1,\v {k}}^{\dag }\psi_{1,\v {k}}
+\psi_{1,-\v {k}}\psi_{1,-\v {k}}^\dag
+\psi_{2,\v {k}}^{\dag }\psi_{2,\v {k }}
+\psi_{2,-\v {k}}\psi_{2,-\v {k}}^\dag
}
\nonumber \\
&=&
(-)^{
\alpha_{\v {k}}^{\dag }\alpha_{\v {k}}
+\beta_{\v {k} }^{\dag }\beta_{\v {k}}
+\alpha_{-\v {k}}\alpha_{-\v {k }}^\dag
+\beta_{-\v {k}}\beta_{-\v {k}}^\dag
}
\nonumber \\
&=&
(-)^{
\alpha_{\v {k}}^{\dag }\alpha_{\v {k}}
+\beta_{\v {k} }^{\dag }\beta_{\v {k}}
+\alpha_{-\v {k}}^\dag\alpha_{-\v {k}}
+\beta_{-\v {k}}^\dag\beta_{-\v {k}}
}
.
\end{eqnarray}
%
%
Hence we have
\begin{eqnarray}
&&(-)^{
\sum\limits_{a=1,2}(\psi_{a,\v {k}}^{\dag }\psi_{a,\v {k} }
+\psi_{a,-\v {k}}^{\dag }\psi_{a,-\v {k}})
}
\mid \Psi_{\text{ mean}}^{(u_{\v i\v j},\eta_{\v i\v j})}\rangle \>
\nonumber \\
&=&
(-)^{
\alpha_{\v {k}}^{\dag }\alpha_{\v {k}}
+\beta_{\v {k} }^{\dag }\beta_{\v {k}}
+\alpha_{-\v {k}}^\dag\alpha_{-\v {k}}
+\beta_{-\v {k}}^\dag\beta_{-\v {k}}
}
\mid \Psi_\text{mean}^{(u_{\v i\v j},\eta _{\v i\v j})}\rangle
\nonumber\\
&=&\mid \Psi_\text{mean}^{(u_{\v i\v j},\eta_{\v i\v j})}\rangle
\end{eqnarray}
for $\v {k}>0$. We see that the total number of the $\psi $ fermions on all the
$\v {k}\neq0$ orbitals is always even.

So to determine if the mean-field ground state contain even or odd
number of $\psi $ fermions, we only need to count the number $\psi$
fermions at the four special points: $(0,0)$, $(0,\pi )$, $(\pi ,0)$,
$(\pi ,\pi )$. For $\v {k}=0$,
\begin{align}
&\ \ \
(-)^{\psi_{1,\v k}^\dag \psi_{1,\v k}
+\psi_{2,\v k}^\dag \psi_{2,\v k}}
=(-)^{
\psi_{1,\v k}^\dag \psi_{1,\v k}
-\psi_{2,\v k}^\dag \psi_{2,\v k} }
\nonumber\\
&=(-)^{ \frac12 \Psi^\dag_{\v k} \si_3\otimes \si_3 \Psi_{\v k} }.
\end{align}
Using \eqn{W33W}, we find
\begin{align}
&\ \ \
\frac12 \Psi^\dag_{\v k} \si_3\otimes \si_3 \Psi_{\v k}
\nonumber\\
&= \frac12
a(\v k)
(
\al_{\v k}^\dag \al_{\v k}
-\al_{\v k} \al_{\v k}^\dag
-\bt_{\v k}^\dag \bt_{\v k}
+\bt_{\v k} \bt_{\v k}^\dag
)
\nonumber\\
&\ \
+\frac12
b(\v k)
(
\al_{\v k}^\dag \bt_{\v k}
-\al_{\v k} \bt_{\v k}^\dag
+\bt_{\v k}^\dag \al_{\v k}
-\bt_{\v k} \al_{\v k}^\dag
)
\nonumber\\
&\ \
+i\frac12
c(\v k)
(
-\al_{\v k}^\dag \bt_{\v k}
-\al_{\v k} \bt_{\v k}^\dag
+\bt_{\v k}^\dag \al_{\v k}
+\bt_{\v k} \al_{\v k}^\dag
)
\nonumber\\
&=
a(\v k) ( \al_{\v k}^\dag \al_{\v k} -\bt_{\v k}^\dag \bt_{\v k})
+b(\v k) ( \al_{\v k}^\dag \bt_{\v k} +\bt_{\v k}^\dag \al_{\v k})
\nonumber\\
&\ \ \ \ \ \ \ \ \ \
+ic(\v k) ( -\al_{\v k}^\dag \bt_{\v k} +\bt_{\v k}^\dag \al_{\v k}) .
\end{align}
We see that
\begin{align}
&\ \ \ (-)^{\psi_{1,\v k}^\dag \psi_{1,\v k}
+\psi_{2,\v k}^\dag \psi_{2,\v k}}
\\
&=e^{\pi i
( a(\v k) ( \al_{\v k}^\dag \al_{\v k} -\bt_{\v k}^\dag \bt_{\v k})
+b(\v k) ( \al_{\v k}^\dag \bt_{\v k} +\bt_{\v k}^\dag \al_{\v k})
+ic(\v k) ( -\al_{\v k}^\dag \bt_{\v k} +\bt_{\v k}^\dag \al_{\v k})
 )
 }
\nonumber .\end{align} If we treat $(\al_{\v k},\bt_{\v k})$ as an
iso-spin-1/2 doublet, then the above operator generates a $2\pi$
rotation since $ a^2(\v k) +b^2(\v k) +c^2(\v k) = 1 $. This is
consistent with the following relation
\begin{align}
& \al_{\v k} (-)^{\psi_{1,\v k}^\dag \psi_{1,\v k} +\psi_{2,\v
k}^\dag \psi_{2,\v k}}
=
-(-)^{\psi_{1,\v k}^\dag \psi_{1,\v k} +\psi_{2,\v k}^\dag \psi
_{2,\v k}} \al_{\v k}
\nonumber\\
& \bt_{\v k} (-)^{\psi_{1,\v k}^\dag \psi_{1,\v k} +\psi_{2,\v
k}^\dag \psi_{2,\v k}}
=
-(-)^{\psi_{1,\v k}^\dag \psi_{1,\v k} +\psi_{2,\v k}^\dag \psi
_{2,\v k}} \bt_{\v k}
\end{align}
Since the operator $(-)^{\al_{\v k}^\dag \al_{\v k} +\bt_{\v k}^\dag
\bt_{\v k}}$ has the same algebra as $(-)^{\psi_{1,\v k}^\dag \psi
_{1,\v k} +\psi_{2,\v k}^\dag \psi_{2,\v k}}$ and the two operators
are equal when $a(\v k)=1$, $b(\v k)=c(\v k)=0$,
we have
\begin{equation}
(-)^{\psi_{1,\v k}^\dag \psi_{1,\v k} +\psi_{2,\v k}^\dag \psi_{2,\v
k}} = (-)^{\al_{\v k}^\dag \al_{\v k} +\bt_{\v k}^\dag \bt_{\v k}}.
\end{equation}
Thus, the total fermion number at the four points, $\v {k} =(0,0)$,
$(0,\pi )$, $(\pi ,0)$, $(\pi ,\pi )$ satisfies
\begin{equation}
(-)^{N_{\v {k}=\v {0}}}=(-)^{\sum\limits_{\v {k=}0}(\alpha_{\v {k}
}^{\dag }\alpha_{\v {k}}+\beta_{\v {k}}^{\dag }\beta_{\v {k}
})}
.
\end{equation}
For a given point of $\v {k}=0,$ the energies for the particles $\alpha $
and $\beta $ are $\varepsilon_{++}(\v {k})$ and $
\varepsilon_{+-}(\v {k})$.  There are two situations:
\begin{enumerate}
\item  For $\varepsilon_{+-}(\v {k})>0$,
the $\v k$-orbital will be filled 0 particle: zero $
\alpha $ particle and zero $\beta $ particle.
\item  For $\varepsilon_{+-}(\v {k})<0$,
the $\v k$-orbital will be filled 1 particle: zero $
\alpha $ particle and one $\beta $ particle.
\end{enumerate}
We find that the total number of $\psi$ fermion at the $\v
k=0$ points and at all $\v k$ are given by
\begin{equation}
\label{Ne2} N_{\v {k}=\v {0}} \text{ mod }2 =N \text{ mod } 2
=\sum\limits_{\v {k=}0} \zeta_{\v k} \text{ mod } 2.
\end{equation}
Note that on even-by-even lattice (ee), all the
four $\v k=0$ points $\v {k}=(0,0)$, $(0,\pi )$, $(\pi ,0)$, $(\pi
,\pi )$ are allowed.  In this case $N$ mod 2 is the sum of all four
$\zeta_{\v k=0}$ mod 2.  On even-by-odd lattice (eo), only two $\v k=0$
points $\v {k}=(0,0)$, $(\pi ,0)$ are allowed.  In this case $N
\text{ mod } 2 = \zeta_{(0,0)}+ \zeta_{(\pi,0)}$ mod 2.


Let's use the topological order 1000 as an example to demonstrate a
detailed calculation of the ground state degeneracy. There are four
degenerate mean-field ground states
$|m,n\>=|\Psi_\text{mean}^{(u_{\v i\v j}^{(m,n)}, \eta_{\v i\v
j}^{(m,n)})}\rangle$, $m,n=0,1$.  On an even-by-even (ee) lattice,
the state $ |0,0\rangle $ has periodic boundary conditions along
both $x$ and $y$ directions. Among the four $\v {k}=0$ points, only
$\v {k}=(0,0)$ point has $\zeta_{\v k}=1$ as indicated by the first
1 in the label 1000.  As a result, the ground state $|0,0\rangle $
contains an odd number of fermions and is un-physical:
$\mathcal{P}|0,0\rangle =0 $.  For the states $|0,1\rangle ,$
$|1,0\rangle ,$ $|1,1\rangle $, the $\v k=(0,0)$ point is not
allowed.  Consequently, $N=0 \text{ mod } 2$.  Hence $|0,1\rangle ,$
$ |1,0\rangle ,$ $|1,1\rangle $ are all physical states.  This gave
rise to three degenerate ground states for the topological order
$\text{1000}$ on an (ee) lattice.

Second, we calculate the ground state degeneracy on an even-by-odd
(eo) [or odd-by-even (oe)] lattice. For the state $|0,0\rangle $, only
two $\v k=0$ points are allowed: $\v {k}=( 0,0) $ and $\v {k} =(\pi
,0)$. Due to $\zeta_{(0,0)}=1$, $|0,0\rangle $ is forbidden, $
\mathcal{P}|0,0\rangle =0$. For other states $|0,1\rangle$,
$|1,0\rangle ,$ $|1,1\rangle $, $\v {k}=(0,0) $ is not allowed on an
(eo) [or (oe)] lattice. For the same reason, one obtains three
degenerate ground states $|0,1\rangle$, $|1,0\rangle$, $|1,1\rangle $
on an (eo) [or (oe)] lattice. Third, we calculate the ground state
degeneracy on an odd-by-odd (oo) lattice. For the state $|0,0\rangle
$, there is only one $\v k=0$ point: $\v {k}=(0,0) $.  As a result,
$|0,0\rangle $ is not permitted by the projection operator,
$\mathcal{P}|0,0\rangle =0$.  However, without the point $\v {k}=(0,0)
$, other states $ |0,1\rangle$, $|1,0\rangle$, $|1,1\rangle $ are
all physical. One also obtains three degenerate ground states
$|0,1\rangle$, $|1,0\rangle$, $ |1,1\rangle$ on an (oo) lattice.

By this method, we obtain topological degeneracies on different lattices for
the $16$ topological orders. The results are given in the following table:
\begin{widetext}
\[
\begin{array}{rcccccccccccccccc}
      &1111 &1110 &1101 &1011 &0111 &1100 &0011 &1001 &0110 &1010 &0101 &1000 &0100 &0010 &0001 &0000 \\
(ee) & 4   & 3   & 3   & 3   & 3   & 4   & 4   & 4   & 4   & 4   & 4   & 3   & 3   & 3   & 3 & 4 \\
(eo) & 4   & 3   & 3   & 3   & 3   & 4   & 4   & 2   & 2   & 2   & 2   & 3   & 3   & 3   & 3 & 4 \\
(oe) & 4   & 3   & 3   & 3   & 3   & 2   & 2   & 2   & 2   & 4   & 4   & 3   & 3   & 3   & 3 & 4 \\
(oo) & \_  & 1   & 1   & 1   & 1   & 2   & 2   & 2   & 2   & 2   & 2   & 3   & 3   & 3   & 3 &
4
\end{array}
\]
\end{widetext}

\subsection{The crystal momentum}

Next, we calculate the crystal momentum for $16$ topological ordered
states. Because the spin Hamiltonian is translation invariance, the
ground states carry definite crystal momentum. To calculate the
crystal momentum $\v {K }$, we note that the fermion wave function
satisfies the (anti) periodic boundary condition.

The crystal momentum is given by
\begin{align}
&\ \ \ \v {\hat{K}}\mid \Psi_{\text{\text{spin}}}\rangle =\sum\limits_{
\v {k}}\v {k}\psi_{\v {k}}^{\dag }\psi_{\v {k}}
\mid \Psi_{\text{\text{spin}}}\rangle
\nonumber\\
&=\sum\limits_{\v k\neq 0}
\v {k}\psi_{\v {k}}^{\dag }\psi_{\v {k}}\mid \Psi_{
\text{\text{spin}}}\rangle +\sum\limits_{\v {k=0}}\v {k}\psi
_{\v {k}}^{\dag }\psi_{\v {k}}\mid \Psi_{\text{\text{spin}}
}\rangle .
\end{align}
The ground state $|\Psi_\text{mean}^{(u_{\v i\v j}^{(m,n)},\eta
_{\v i\v j}^{(m,n)})}\rangle $ at $\v k \neq 0$ has a form $
\prod\limits_{\v {k>}0}{\alpha_{\v {k}}\alpha_{-\v {k}}\beta_{
\v {k}}\beta_{-\v {k}}}|0\rangle_\psi $
where $|0\rangle_\psi$ is the state with no $\psi$ fermion.
Thus, the total crystal momentum $\v {K}$ are obtained as
\begin{align}
&\ \ \ \v {\hat{K}}\mid \Psi_{\text{\text{spin}}}\rangle
=
\sum\limits_{\v {k=0}}\v {k}\psi
_{\v {k}}^{\dag }\psi_{\v {k}}\mid \Psi_{\text{\text{spin}}
}\rangle
\nonumber\\
&=
\sum_{\v k=0} \v k \zeta_{\v k}
\end{align}
where we have used (at $\v k=0$)
\begin{align}
 \psi_{\v k}^\dag\psi_{\v {k}} \text{ mod } 2 = \zeta_{\v k} .
\end{align}
Thus, to determine the crystal momentum, we only need to focus on the
cases at $ \v {k}=0$.

By this method, we obtain the crystal momenta on different lattices for $16$
topological orders. The following tables show the crystal momenta of
different ground states:
\[
\begin{array}{lllll}
\v {K}\text{ }(0000) & (ee) & (eo) & (oe) & (oo) \\
(00) & (0,0) & (0,0) & (0,0) & (0,0) \\
(01) & (0,0) & (0,0) & (0,0) & (0,0) \\
(10) & (0,0) & (0,0) & (0,0) & (0,0) \\
(11) & (0,0) & (0,0) & (0,0) & (0,0)
\end{array}
\]
\[
\begin{array}{lllll}
\v {K}\text{ }(0011) & (ee) & (eo) & (oe) & (oo) \\
(00) & (0,\pi ) & - & (0,0) & (0,0) \\
(01) & (0,0) & - & (0,0) & (0,0) \\
(10) & (0,0) & (0,0) & (0,\pi ) & - \\
(11) & (0,0) & (0,0) & (0,0) & -
\end{array}
\]
\[
\begin{array}{lllll}
\v {K}\text{ }(1100) & (ee) & (eo) & (oe) & (oo) \\
(00) & (0,\pi ) & - & (0,\pi ) & - \\
(01) & (0,0) & - & (0,0) & - \\
(10) & (0,0) & (0,0) & (0,0) & (0,0) \\
(11) & (0,0) & (0,0) & (0,0) & (0,0)
\end{array}
\]
\[
\begin{array}{lllll}
\v {K}\text{ }(1111) & (ee) & (eo) & (oe) & (oo) \\
(00) & (0,0) & (\pi ,0) & (0,\pi ) & - \\
(01) & (0,0) & (\pi ,0) & (0,0) & - \\
(10) & (0,0) & (0,0) & (0,\pi ) & - \\
(11) & (0,0) & (0,0) & (0,0) & -
\end{array}
\]
\[
\begin{array}{lllll}
\v {K}\text{ }(0101) & (ee) & (eo) & (oe) & (oo) \\
(00) & (\pi ,0) & (0,0) & - & (0,0) \\
(01) & (0,0) & (\pi ,0) & (0,0) & - \\
(10) & (0,0) & (0,0) & - & (0,0) \\
(11) & (0,0) & (0,0) & (0,0) & -
\end{array}
\]
\[
\begin{array}{lllll}
\v {K}\text{ }(1010) & (ee) & (eo) & (oe) & (oo) \\
(00) & (\pi ,0) & (\pi ,0) & - & - \\
(01) & (0,0) & (0,0) & (0,0) & (0,0) \\
(10) & (0,0) & (0,0) & - & - \\
(11) & (0,0) & (0,0) & (0,0) & (0,0)
\end{array}
\]
\[
\begin{array}{lllll}
\v {K}\text{ }(0110) & (ee) & (eo) & (oe) & (oo) \\
(00) & (\pi ,\pi ) & - & - & (0,0) \\
(01) & (0,0) & - & (0,0) & - \\
(10) & (0,0) & (0,0) & - & - \\
(11) & (0,0) & (0,0) & (0,0) & (0,0)
\end{array}
\]
\[
\begin{array}{lllll}
\v {K}\text{ }(1001) & (ee) & (eo) & (oe) & (oo) \\
(00) & (\pi ,\pi ) & - & - & - \\
(01) & (0,0) & - & (0,0) & (0,0) \\
(10) & (0,0) & (0,0) & - & (0,0) \\
(11) & (0,0) & (0,0) & (0,0) & -
\end{array}
\]
\[
\begin{array}{lllll}
\v {K}\text{ }(1000) & (ee) & (eo) & (oe) & (oo) \\
(00) & - & - & - & - \\
(01) & (0,0) & (0,0) & (0,0) & (0,0) \\
(10) & (0,0) & (0,0) & (0,0) & (0,0) \\
(11) & (0,0) & (0,0) & (0,0) & (0,0)
\end{array}
\]
\[
\begin{array}{lllll}
\v {K}\text{ }(0100) & (ee) & (eo) & (oe) & (oo) \\
(00) & - & (0,0) & - & (0,0) \\
(01) & (0,0) & - & (0,0) & - \\
(10) & (0,0) & (0,0) & (0,0) & (0,0) \\
(11) & (0,0) & (0,0) & (0,0) & (0,0)
\end{array}
\]
\[
\begin{array}{lllll}
\v {K}\text{ }(0010) & (ee) & (eo) & (oe) & (oo) \\
(00) & - & - & (0,0) & (0,0) \\
(01) & (0,0) & (0,0) & (0,0) & (0,0) \\
(10) & (0,0) & (0,0) & - & - \\
(11) & (0,0) & (0,0) & (0,0) & (0,0)
\end{array}
\]
\[
\begin{array}{lllll}
\v {K}\text{ }(0001) & (ee) & (eo) & (oe) & (oo) \\
(00) & - & (0,0) & (0,0) & (0,0) \\
(01) & (0,0) & - & (0,0) & (0,0) \\
(10) & (0,0) & (0,0) & - & (0,0) \\
(11) & (0,0) & (0,0) & (0,0) & -
\end{array}
\]
\[
\begin{array}{lllll}
\v {K}\text{ }(0111) & (ee) & (eo) & (oe) & (oo) \\
(00) & - & - & - & (0,0) \\
(01) & (0,0) & (\pi ,0) & (0,0) & - \\
(10) & (0,0) & (0,0) & (0,\pi ) & - \\
(11) & (0,0) & (0,0) & (0,0) & -
\end{array}
\]
\[
\begin{array}{lllll}
\v {K}\text{ }(1011) & (ee) & (eo) & (oe) & (oo) \\
(00) & - & (\pi ,0) & - & - \\
(01) & (0,0) & - & (0,0) & (0,0) \\
(10) & (0,0) & (0,0) & (0,\pi ) & - \\
(11) & (0,0) & (0,0) & (0,0) & -
\end{array}
\]
\[
\begin{array}{lllll}
\v {K}\text{ }(1101) & (ee) & (eo) & (oe) & (oo) \\
(00) & - & - & (0,\pi ) & - \\
(01) & (0,0) & (\pi ,0) & (0,0) & - \\
(10) & (0,0) & (0,0) & - & (0,0) \\
(11) & (0,0) & (0,0) & (0,0) & -
\end{array}
\]
\[
\begin{array}{lllll}
\v {K}\text{ }(1110) & (ee) & (eo) & (oe) & (oo) \\
(00) & - & (\pi ,0) & (0,\pi ) & - \\
(01) & (0,0) & - & (0,0) & - \\
(10) & (0,0) & (0,0) & - & - \\
(11) & (0,0) & (0,0) & (0,0) & (0,0)
\end{array}
.
\]

\section{How many distinct $Z_2$ topological orders?}

We would like to point out that the four topological indices
$\zeta_{\v k}$ at $\v k=0$ really describe 16 classes of mean-field
ansatz. It is not clear if different mean-field ansatz give rise to
different many-body spin wave functions.  So it is possible that the
16 sets of topological indices $\zeta_{\v k}$ describe less than 16
class of $Z_2$ topological orders.

On the other hand, if two $Z_2$ topological phases
can be separated through measurable physical quantities,
such as ground state degeneracy and crystal momenta,
then the two topological phases will be really distinct.

Using the ground state degeneracies on different types of lattice we
can group the 16 sets of topological indices $\zeta_{\v k}$ into 7
groups: $\{0000\}$, $\{ 0001, 0010, 0100, 1000 \}$, $\{ 0101, 1010
\}$, $\{ 1100, 0011 \}$, $\{ 1001, 0110 \}$, $\{ 0111, 1011, 1101,
1110 \}$, $\{ 1111 \}$. Each group have the same ground state
degeneracies.  So we have at least 7 distinct $Z_2$ topological
phases.  If we assume that the boundary condition labels $(m,n)$ are
not physically observable, the crystal momentum distributions cannot
further separate the above 7 groups into smaller groups.  However,
it is likely that the boundary condition labels $(m,n)$ are
physically observable by moving the unique type of fermionic
excitations (the spinons) around the torus. In this case all 16 sets
of topological indices $\zeta_{\v k}$ label distinct $Z_2$
topological phases.

\section{Two examples of Z2A topological orders}

Let us discuss two different translation symmetric Z2A topological
orders studied in \cite{KLW0800} in more detail.  The first one is
described by the following ``mean-field'' fermion
Hamiltonian\cite{Wqoslpub} within the projective construction
\cite{Wsrvb}:
\begin{eqnarray}
\label{Hmean11}
H_\text{mean} &=&\sum_{\v i\v j}\psi_{\v i}^{\dag }u_{\v i\v j}\psi_{\v j}+\sum_{\v
i}\psi
_{\v i}^{\dag }a_{\v i}\psi_{\v i}  \label{1} \\
u_{i,i+x} &=&u_{i,i+\v{y}}=-\chi \sigma ^3,  \nonumber \\
u_{i,i+x+y} &=&\eta \sigma ^1+\lambda \sigma ^2,  \nonumber \\
u_{i,i-x+y} &=&\eta \sigma ^1-\lambda \sigma ^2,  \nonumber \\
a_{\v i}^1 &=&\upsilon .  \nonumber
\end{eqnarray}
where $\psi ^T=(\psi_1,\psi_2)$.  The other Z2A spin liquid comes from
an exact soluble spin-1/2 model on square lattice - the Wen-plaquette
model.\cite{Wqoexct} Its Hamiltonian is $ H=16g\sum_{\v i}S_{\v
i}^yS_{\v i+\v{x}}^xS_{\v i+\v{x}+\v{y}}S_{\v i+\v{y}}^x$.  Using
projective construction, one can introduce a ``mean-field'' fermion
Hamiltonian:\cite{Wqoslpub} to describe such a spin liquid:
\begin{equation}
H_\text{mean}=\sum_{\<\v i\v j\> }\left( \psi_{I,i}^{\dag }u_{\v
i\v j}^{IJ}\psi
_{J,j}+\psi_{I,i}^{\dag }\eta_{\v i\v j}^{IJ}\psi_{J,j}^{\dag }+h.c.\right)
\label{Hmean}
\end{equation}
where $I,J=1,2$. It is known that the ground states for $g<0$ and
$g>0$ have different symmetry protected topological orders.  The
ground state (Z2A topological order) for $g<0$ is described by
mean-field ansatz $-\eta _{i,i+x}=u_{i,i+x}=1+\sigma ^3$ and
$-\eta_{i,i+y}=u_{i,i+y}=1-\sigma ^3$.  The ground state (Z2B
topological order) for $g>0$ is described by mean-field ansatz
$-\eta_{i,i+x}=u_{i,i+x}=(-)^{i_y}(1+\sigma ^3)$ and $-\eta
_{i,i+y}=u_{i,i+y}=1-\sigma ^3$.

We like to ask: whether the topological order for Z2A gapped state in
Eq. (\ref{Hmean11}) and the topological order for the Wen-plaquette
model in Eq.(\ref{Hmean}) are the same one.  The four $Z_2$
topological variables $\zeta_{\v k=0} $ can help us to answer the
question.

For the Z2A gapped state, a ``mean-field'' fermion Hamiltonian in momentum
space becomes
\begin{equation}
H_\text{mean}(\v {k})=\sum_{\v {k}\>}\Psi_{\v {k}}^{\dag }U(
\v {k})\Psi_{\v {k}}+h.c.
\end{equation}
where
\begin{eqnarray*}
U(\v {k}) &=&\sum_{\{\alpha ,\beta \}}c_{\{\alpha \beta \}}(\v {k}
)M_{\{\al\beta \}}, \\
c_{\{33\}} &=&\cos k_x+\cos k_y, \\
c_{\{31\}} &=&\eta (\cos (k_x+k_y)+\cos (k_x-k_y))+\upsilon , \\
c_{\{02\}} &=&\lambda (\cos (k_x+k_y)-\cos (k_x-k_y)).
\end{eqnarray*}
Because $M_{\{ 33\} },$ $M_{\{ 31\} }$and $M_{\{
02\} }$ make up of an anticommutation base, we have
\begin{widetext}
\begin{eqnarray*}
&&\varepsilon_{+-}(\v {k} ) =0-\sqrt{c_{\{33\}}^2(\v {k}
)+c_{\{31\}}^2(\v {k})+c_{\{02\}}^2(\v {k})} \\
&=&-\sqrt{\left( \cos k_x+\cos k_y\right) ^2+\left[ \eta (\cos (k_x+k_y)+\cos
(k_x-k_y))+\upsilon \right] ^2+\left[ \lambda (\cos (k_x+k_y)-\cos
(k_x-k_y))\right] ^2}
\end{eqnarray*}
\end{widetext}
For the Z2A gapped state, one has
\[
\zeta_{(0,0)}=1,\ \
\zeta_{(0,\pi)}=1, \ \
\zeta_{(\pi,0)}=1, \ \
\zeta_{(\pi,\pi)}=1.
\]
It belongs to the $\text{1111}$ type of the topological order. The
topological degeneracy is $4,$ $4,$ $4,$ for even-by-even,
even-by-odd, odd-by-even
respectively. On
odd-by-odd lattices, the state has no energy gap.

Another example of topological order is the Wen-plaquette model.
The ``mean-field'' ansatz in
momentum space now becomes
\begin{eqnarray*}
U(\v {k}) &=&\sum_{\{\alpha ,\beta \}}c_{\{\alpha\beta \}}(\v {k}
)M_{\{\al\beta \}}, \\
c_{\{30\}} &=&\cos k_x+\cos k_y, \\
c_{\{33\}} &=&\cos k_x-\cos k_y, \\
c_{\{20\}} &=&\sin k_x+\sin k_y, \\
c_{\{23\}} &=&\sin k_x-\sin k_y.
\end{eqnarray*}
There are two even matrices in the ansatz, $M_{\{ 30\} }$ and $
M_{\{ 33\} }$. We have
\[
\varepsilon_{+-}(\v {k}=0)=\left| \cos k_x+\cos k_y\right| -\left|
\cos k_x-\cos k_y\right| .
\]
For the topological order of the Wen-plaquette model, one has
\[
\zeta_{(0,0)}=0,\ \ \zeta_{(0,\pi)}=1, \ \ \zeta_{(\pi,0)}=1,\ \
\zeta_{(\pi,\pi)}=0.
\]
It belongs to the \text{0110} type of the topological order. The
topological degeneracy is $4$, $2$, $2$, $2$ for $(ee)$, $(eo)$,
$(oe)$, $(oo)$ lattices. Because topological order in the toric-code
model with translation invariance\cite{K032} and that in the
Wen-plaquette model are equivalent, one can use the same
wave-function to describe the toric code model.

Then when one changes the Z2A gapped state denoted by $\text{1111},$
into the topological order of the Wen-plaquette model denoted by
\text{0110}$,$ quantum phase transition occurs with emergent
massless fermion at $\v {k} =(0,0)$ and $\v {k}=(\pi ,\pi )$.

\section{Non-Abelian topological states}

We like to point out that for the four Z2A topological states
described by $\{\zeta_{\v k}\}=$ 1000, 0100, 0010, 0001, the
corresponding mean-field Hamiltonian describes a superconducting
state whose band structure has an odd winding
number.\cite{SMF9945,RG0067} So those four Z2A topological states
are closely related to the topological spin liquid state obtained by
the projection of $p_x+i p_y$ SC states.\cite{RG0067} As a result,
the four Z2A topological states have a topological order described
by Ising topological quantum field theory (which is the same
topological quantum field theory describing the non-Abelian Paffien
FQH state at $\nu=1/2$\cite{MR9162,Wnabhalf,NW9629,Wpcon,NSS0883}).
This is consistant with the fact that those four Z2A topological
states all have 3 degenerate ground states on torus and do not have
time reversal symmetry.  Therefore the four Z2A topological states
are non-Abelian states with excitations that carry non-Abelian
statistics described by Ising topological quantum field theory.

We believe that the four Z2A topological states described by
$\{\zeta_{\v k}\}=$ 0111, 1011,1101, 1110 are also non-Abelian
states described by Ising topological quantum field theory.
However, 0111, 1011,1101, 1110 states are different from the 1000,
0100, 0010, 0001 states since thay have different ground state
degeneracy on (oo) lattices.  Some concrete constructions of those
non-Abelian topological states and other topological states
discussed in this paper are given in appendix \ref{const}.

We also like to point out that the mean-field Hamiltonian is a
Hamiltonian for a superconductor.  The $Z_2$ topological indices
$\zeta_{\v k}$ provide a classification of translation invariant 2D
topological superconductors as discussed in \Ref{KWtopSC}.

\section{Conclusion}

In this paper, we study topological phases that have the translation
symmetry using the projective approach. We concentrated on a class
topological phases described by Z2A PSG.  We find that 2D Z2A
topological phases can be further divided into 16 classes, which can
be described by four $Z_2$ topological variables $\zeta_{\v k}=0,1$
at the four $\v k=$ points. Through the projected SC wave-functions,
we obtain the topological degeneracies and the crystal momentum for
those 16 classes of topological states.  This allows us to identify
the topological phase of the Wen's plaquette model as the 0110 Z2A
topological order. In addition, it is predicted that massless
fermionic excitations appear at the quantum phase transition between
different topological orders with translation invariance.

This research is supported by NSF Grant No. DMR-0706078, NFSC no. 10228408,
and NFSC no. 10874017.

\appendix

\section{Definition of anti-commutation frames}

In this appendix we define anti-commutation bases. An anti-commutation base $
(M_{\{ \alpha \beta \} },M_{\{ \alpha ^{\prime }\beta
^{\prime }\} },M_{\{ \alpha ^{\prime \prime }\beta ^{\prime
\prime }\} },...)$ is a maximum set for several $4\times 4$ matrices
which anti-commute each other
\begin{eqnarray*}
\{ M_{\{ \alpha \beta \} },M_{\{ \alpha ^{\prime }\beta
^{\prime }\} }\} &=&0, \\
\{ M_{\{ \alpha ^{\prime }\beta ^{\prime }\} },M_{\{
\alpha ^{\prime \prime }\beta ^{\prime \prime }\} }\} &=&0, \\
\{ M_{\{ \alpha \beta \} },M_{\{ \alpha ^{\prime \prime
}\beta ^{\prime \prime }\} }\} &=&0, \\
&&...
\end{eqnarray*}
If a matrix can be decomposed into an anti-commutation base,
\[
U=\sum_{\{\alpha ,\beta \}}c_{\{\alpha\beta \}}M_{\{\al\beta \}}
\]
one has the determine of the matrix $U$ as
\[
\det U=[c_{\{\alpha\beta \}}^2+c_{\{\alpha ^{\prime }\beta ^{\prime
}\}}^2+c_{\{\alpha ^{\prime \prime }\beta ^{\prime \prime }\}}^2+...]^2.
\]
Any matrix out of the base $M_{\{ \tilde{\alpha}\tilde{\beta}\} }$
can be considered as a particular direction in the base, $M_{\{ \tilde{
\alpha}\tilde{\beta}\} }=\sum_{\{\alpha ,\beta \}}\tilde{c}_{\{\alpha
\beta \}}M_{\{\al\beta \}}.$ When one add it to $U$, we have the
determine as
\begin{align}
\det U&=[(c_{\{\alpha \beta \}}+\tilde{c}_{\{\alpha \beta
\}})^2+(c_{\{\alpha ^{\prime }\beta ^{\prime }\}}+\tilde{c}_{\{\alpha
^{\prime }\beta ^{\prime }\}})
\nonumber\\
&\ \ \ \ \ \ \ \ \ \
+(c_{\{\alpha ^{\prime \prime }\beta ^{\prime \prime }\}}
+\tilde{c}_{\{\alpha ^{\prime \prime }\beta ^{\prime
\prime }\}})^2+...]^2.
\end{align}

A famous example is five $\gamma $ matrices, $M_{\{ 33\} },$ $
M_{\{ 13\} },$ $M_{\{ 11\} }$, $M_{\{ 02\}
} $ and $M_{\{ 01\} }$ making up of an anti-commutation base. If
the matrix has the form as
\[
U=\sum\limits_{\{\alpha ,\beta \}}c_{\{\alpha \beta \}}M_{\{\al\beta
\}},\text{ }\{\al\beta \}=\{ 33\} ,\{ 13\}
,\{ 11\} ,\{02\},\{01\},
\]
one has the eigenvalue of $U$ as
\[
\pm \sqrt{c_{\{33\}}^2+c_{\{13\}}^2+c_{\{11\}}^2+c_{\{02\}}^2+c_{\{01\}}^2}
\]
and the determine of $U$ as
\[
\det U=\left(
c_{\{33\}}^2+c_{\{13\}}^2+c_{\{11\}}^2+c_{\{02\}}^2+c_{\{01\}}^2\right) ^2.
\]

$M_{\{ 30\} },$ $M_{\{ 12\} }$ and $M_{\{
22\} }$ make up of an anti-commutation base, I. If there is a matrix
as
\[
U=\sum_{\{\alpha ,\beta \}}c_{\{\alpha \beta \}}M_{\{\al\beta \}},
\text{ }\{\al\beta \}=\{ 30\} ,\{ 12\} ,\{
22\} .
\]
One has the eigenvalue of $U$ as
\[
\pm \sqrt{c_{\{30\}}^2+c_{\{12\}}^2+c_{\{22\}}^2}
\]
and diagonalize $U=c_{\{3,0\}}M_{\{3,0\}}+c_{\{1,2\}}M_{\{1,2\}}+c_{\{2,2
\}}M_{\{2,2\}}$ into a $4\times 4$ matrix as
\[
\sqrt{c_{\{30\}}^2+c_{\{12\}}^2+c_{\{22\}}^2}\cdot \left(
\begin{array}{llll}
1 & 0 & 0 & 0 \\
0 & -1 & 0 & 0 \\
0 & 0 & 1 & 0 \\
0 & 0 & 0 & -1
\end{array}
\right) .
\]
$M_{\{ 33\} },$ $M_{\{ 31\} }$ and $M_{\{
02\} }$ make up of another anti-commutation base, II. One can
diagonalize $U=c_{\{3,3\}}M_{\{3,3\}}+c_{\{3,1\}}M_{\{3,1\}}+c_{\{0,2\}}M_{
\{0,2\}},$ into another $4\times 4$ matrix as
\[
\sqrt{c_{\{33\}}^2+c_{\{31\}}^2+c_{\{02\}}^2}\cdot \left(
\begin{array}{llll}
1 & 0 & 0 & 0 \\
0 & -1 & 0 & 0 \\
0 & 0 & -1 & 0 \\
0 & 0 & 0 & 1
\end{array}
\right) .
\]

\section{The examples of the ansatzs of different Z2A topological
orders}
\label{const}

In this part we give one example for each type of topological order. The
ansatzs $U(\v {k})$ in momentum space for the 16 classes topological orders
are given by
\begin{align*}
0000: & \left(
\begin{array}{cc}
C_1 & 0 \\
0 & C_1
\end{array}
\right)
,&
0011: & \left(
\begin{array}{cc}
\sigma ^3 & 0 \\
0 & B_x
\end{array}
\right)
\end{align*}
\begin{align*}
0101: & \left(
\begin{array}{cc}
\sigma ^3 & 0 \\
0 & B_y
\end{array}
\right)
,&
0110: & \left(
\begin{array}{cc}
\sigma ^3 & 0 \\
0 & B_{x+y}
\end{array}
\right)
\end{align*}
\begin{align*}
1111: & \left(
\begin{array}{cc}
C_1 & 0 \\
0 & -C_1
\end{array}
\right)
, &
1100: & \left(
\begin{array}{cc}
\sigma ^3 & 0 \\
0 & -B_x
\end{array}
\right)
\end{align*}
\begin{align*}
1010: & \left(
\begin{array}{cc}
\sigma ^3 & 0 \\
0 & -B_y
\end{array}
\right)
, &
1001: & \left(
\begin{array}{cc}
\sigma ^3 & 0 \\
0 & -B_{x+y}
\end{array}
\right)
\end{align*}
\begin{align*}
1110: & \left(
\begin{array}{cc}
\sigma ^3 & 0 \\
0 & -C_1
\end{array}
\right)
, &
1011: & \left(
\begin{array}{cc}
\sigma ^3 & 0 \\
0 & -C_2
\end{array}
\right)
\end{align*}
\begin{align*}
0111: & \left(
\begin{array}{cc}
\sigma ^3 & 0 \\
0 & -C_3
\end{array}
\right)
, &
1101: & \left(
\begin{array}{cc}
\sigma ^3 & 0 \\
0 & -C_4
\end{array}
\right)
\end{align*}
\begin{align*}
0001: & \left(
\begin{array}{cc}
\sigma ^3 & 0 \\
0 & C_1
\end{array}
\right)
, &
0100: & \left(
\begin{array}{cc}
\sigma ^3 & 0 \\
0 & C_2
\end{array}
\right)
\end{align*}
\begin{align*}
1000: & \left(
\begin{array}{cc}
\sigma ^3 & 0 \\
0 & C_3
\end{array}
\right)
, &
0010: & \left(
\begin{array}{cc}
\sigma ^3 & 0 \\
0 & C_4
\end{array}
\right)
\end{align*}
The parameters above are defined as
\begin{eqnarray*}
\sigma ^3 &=&\left(
\begin{array}{cc}
1+\frac 14\cos k_x+\frac 14\cos k_y & 0 \\
0 & -1-\frac 14\cos k_x-\frac 14\cos k_y
\end{array}
\right) , \\
B_x &=&\left(
\begin{array}{cc}
\cos k_x & i\sin k_x \\
-i\sin k_x & -\cos k_x
\end{array}
\right) , \\
B_y &=&\left(
\begin{array}{cc}
\cos k_y & i\sin k_y \\
-i\sin k_y & -\cos k_y
\end{array}
\right) , \\
B_{x+y} &=&\left(
\begin{array}{cc}
\cos (k_x+k_y) & i\sin (k_x+k_y) \\
-i\sin (k_x+k_y) & -\cos (k_x+k_y)
\end{array}
\right) ,
\end{eqnarray*}
\begin{widetext}
\begin{eqnarray*}
C_1 &=&\left(
\begin{array}{cc}
\cos k_x-\frac 12\cos (k_x+k_y)+\cos k_y & \sin k_x+i\sin k_y \\
\sin k_x-i\sin k_y & -\cos k_x+\frac 12\cos (k_x+k_y)-\cos k_y
\end{array}
\right) , \\
C_2 &=&\left(
\begin{array}{cc}
\cos k_x-\frac 12\cos (k_x+k_y)-\cos k_y & \sin k_x+i\sin k_y \\
\sin k_x-i\sin k_y & -\cos k_x-\frac 12\cos (k_x+k_y)+\cos k_y
\end{array}
\right) , \\
C_3 &=&\left(
\begin{array}{cc}
\cos k_x+\cos (k_x+k_y)+\frac 12\cos k_y & \sin k_x+i\sin k_y \\
\sin k_x-i\sin k_y & -\cos k_x-\cos (k_x+k_y)-\frac 12\cos k_y
\end{array}
\right) , \\
C_4 &=&\left(
\begin{array}{cc}
\cos k_x+\cos (k_x+k_y)-\frac 12\cos k_y & \sin k_x+i\sin k_y \\
\sin k_x-i\sin k_y & -\cos k_x-\cos (k_x+k_y)+\frac 12\cos k_y
\end{array}
\right) .
\end{eqnarray*}
\end{widetext}
One can check above results by the following table
\[
\begin{array}{lllll}
& (0,0) & (0,\pi ) & (\pi ,0) & (\pi ,\pi ) \\
1+\frac 14\cos k_x+\frac 14\cos k_y & >0 & >0 & >0 & >0 \\
\cos k_x & >0 & >0 & <0 & <0 \\
\cos k_y & >0 & <0 & >0 & <0 \\
\cos (k_x+k_y) & >0 & <0 & <0 & >0 \\
\cos k_x+\cos k_y-\frac 12\cos (k_x+k_y) & >0 & >0 & >0 & <0 \\
\cos k_x-\cos k_y-\frac 12\cos (k_x+k_y) & <0 & >0 & <0 & <0 \\
\cos k_x+\cos (k_x+k_y)+\frac 12\cos k_y & >0 & <0 & <0 & <0 \\
\cos k_x+\cos (k_x+k_y)-\frac 12\cos k_y & >0 & >0 & <0 & >0
\end{array}
.
\]


\end{document}